\documentclass[12pt]{article}

\usepackage{latexsym,amsmath,amssymb,theorem,epsfig}

\usepackage{graphicx}
\usepackage{epstopdf}
\usepackage{amssymb}
\usepackage{amsmath}
\usepackage{psfrag}
\usepackage{epsfig}
\usepackage{cancel}

 \allowdisplaybreaks[4]

\usepackage{color}

\DeclareFontFamily{OT1}{rsfs10}{}
\DeclareFontShape{OT1}{rsfs10}{m}{n}{ <-> rsfs10 }{}
\DeclareMathAlphabet{\mathscript}{OT1}{rsfs10}{m}{n}

\numberwithin{equation}{section}

\newcommand{\bra}{\langle}
\newcommand{\ket}{\rangle}

\newcommand{\x}{{\vec{x}}}

\newcommand{\kk}{{\vec{k}}}

\newcommand{\sfrac}[2]{{\textstyle\frac{#1}{#2}}}
\newcommand\di{\partial}

\def\gsim{ \lower .75ex \hbox{$\sim$} \llap{\raise .27ex \hbox{$>$}} }
\def\lsim{ \lower .75ex \hbox{$\sim$} \llap{\raise .27ex \hbox{$<$}} }
\def\be{\begin{equation}}
\def\ee{\end{equation}}
\def\bea{\begin{eqnarray}}
\def\eea{\end{eqnarray}}

\usepackage{graphicx}

\newcommand{\ba}{\begin{array}}
\newcommand{\ea}{\end{array}}

\begin{document}

%
%
%

\begin{center}
\Large{\textbf{Spontaneous Symmetry Probing}} \\[0.5cm]
 
\large{Alberto Nicolis$^{\rm a}$ and Federico Piazza$^{\rm b}$}
\\[0.5cm]

\small{
\textit{$^{\rm a}$ Department of Physics and ISCAP, \\ Columbia University, New York, NY 10027, USA}}

\vspace{.2cm}

\small{
\textit{$^{\rm b}$ Paris Center for Cosmological Physics and Laboratoire APC,\\  Universit\'e Paris 7, 75205 Paris, France
}}

\end{center}

\vspace{.8cm}

\date{}

\begin{abstract}
For relativistic quantum field theories, we consider Lorentz breaking, spatially homogeneous field configurations or states that evolve in time along a symmetry direction. We dub this situation ``spontaneous symmetry probing" (SSP). We mainly focus on internal symmetries, i.e.~on symmetries that commute with the Poincar\'e group. We prove that the fluctuations around SSP states have a Lagrangian that is explicitly time independent, and we provide the field space parameterization that makes this manifest.
We show that there is always a gapless Goldstone excitation that perturbs the system in the direction of motion in field space.  Perhaps more interestingly, we show that if such a direction is part of a non-Abelian group of symmetries, the Goldstone bosons associated with spontaneously broken generators that do not commute with the SSP one acquire a {\em gap}, proportional to the SSP state's ``speed''. We outline possible applications of this formalism to inflationary cosmology. 
\end{abstract}

%

\section{Introduction}\label{intro}

Symmetry has proven a very powerful tool to deduce general, model independent statements. This is especially true  for quantum field theory, and for its application to particle physics. A celebrated example is the (softly broken) chiral symmetry of strong interactions, that allows to determine the low energy dynamics of pions  regardless of the details of the fundamental theory, QCD. 
In this paper we explore the general implications of symmetries for  \emph{time-dependent} states. In particular, we are interested in field configurations that spontaneously ``probe" the symmetry---or one of the symmetries---in the sense that they evolve in time along a symmetry direction in field space. As we describe at some length in the concluding section of the paper, our main motivation comes from inflationary cosmology, and the systems we consider capture certain essential features of generic early universe inflationary models regardless of their microphysical details. 
But the same formalism applies also to other---apparently disparate---situations, such as many-body systems at finite chemical potential, as we argue below.

We realize that many of our results will sound obvious to some of our readers. Also we claim no originality for some of our results---for instance there is a non-trivial overlap with part of ref.~\cite{ghost}. However we find it useful to characterize the properties we study in broad generality, and beyond the semiclassical approximation, to understand to what extent they follow purely from symmetry considerations. Moreover, some of our results---most notably the `higgsing' of certain would-be Golstone bosons for non-abelian symmetries---were not obvious to us beforehand, and we hope some of our readers will not find them obvious either.

\subsection{Preliminary considerations}
Consider a Poincar\'e invariant field theory in Minkowski space, featuring a continuous symmetry. As a result of Noether's theorem, there is a conserved current operator $J^\mu(x)$, $\partial_\mu  J^\mu(x) = 0$ (Heisenberg picture is understood throughout),
from which we can define a charge operator in the usual way: 
\begin{equation} \label{chargecons}
Q = \int \! d^3 x \,  J^0(\vec{x},t), \qquad \frac{d Q}{dt}=0 \; .
\end{equation}
The operator $Q$ is the symmetry generator for states and operators in the theory. 
We now consider states that spontaneously break both this symmetry and time-translational invariance, but in such a way that time-evolution moves the system along the symmetry direction. We dub this situation ``spontaneous symmetry probing" (SSP).
Denoting by $H$ the Hamiltonian of the system, we therefore simply ask that the system be in a state $|c\ket$ obeying
\begin{equation} \label{ssp2}
\langle c |[H,A(x)] | c \rangle \, = \, c \, \langle c |[Q,A(x)] | c \rangle  \; ,
\end{equation}
for any local operator $A(x)$ and for some parameter $c$. We are simply demanding that any local quantity evolves in time as if acted upon by the symmetry transformation. 

In practice we will often use a slightly looser definition than \eqref{ssp2}. The latter is obviously satisfied if $|c\ket$ is an eigenvector of $H-cQ$, i.e.,
\begin{equation} \label{ssp_lambda}
(H -cQ) |c\ket =  \lambda |c\ket \; ,
\end{equation}
for some real number $\lambda$. 
If the spectrum of this operator is---for a given $c$---bounded from below, we will assume that our $|c \ket$ is the {\em ground state} of this modified Hamiltonian.
Given translational invariance, if $\lambda$ does not vanish, it diverges with the volume. 
From now on we will drop $\lambda$, because we can shift it to zero by redefining the zero of $H$.  We will therefore assume
\be \label{ssp}
(H -cQ) |c\ket = 0 \; ,
\ee
rather than \eqref{ssp_lambda}.
But one should keep in mind that in general $\lambda$ depends on $c$, and therefore different SSP states with different values of $c$ 
have different values of $\lambda$. In this sense $\lambda$ is unlike a cosmological constant, which one can set to zero once for all.

We can interpret eq.~\eqref{ssp} as saying that time translations and the symmetry under consideration are broken by the state $|c\ket$, but there exists an unbroken combination of the two. The same situation arises e.g.~for the ghost condensate \cite{ghost,ghost2}, which is a somewhat degenerate example of SSP.
As for the standard cases of spontaneous symmetry breaking, care should be taken with expressions like \eqref{ssp}. For homogeneous states, both charge and energy are infinite (or zero, in the unbroken case). Therefore, such a relation should be IR-regulated by considering the system at finite volume or, as in \eqref{ssp2}, meant to hold inside matrix elements of commutators with local field operators.   

If the dimension of the symmetry group is higher than one and several of these symmetries are spontaneously broken, eq. \eqref{ssp} is readily generalized by allowing a suitable combination of symmetry generators, 
\begin{equation} \label{sspmany}
\Big(H -  \sum_ a c_a\, Q_a \Big)\, |\vec c \ket = 0 \; .
\end{equation}
For simplicity, we will restrict ourselves to homogeneous states, i.e.~to states that do not break spacial translations.

Looking at things in the Lagrangian formalism allows to be more explicit. 
Let us consider $N$  fields $\phi_n$, $n = 1\dots N$, and say that the infinitesimal transformation  
$\phi_n \rightarrow \phi_n + \epsilon \, \delta \phi_n$  is a symmetry of the theory. This means that the Lagrangian varies by at most a total derivative,
 \begin{equation}
\frac{ \delta {\cal L}}{\delta \phi_n} \delta \phi_n = \partial_\mu K^{\mu}\, . 
\end{equation}
Here
\begin{equation} \label{transformationss}
\delta \phi_n (x) =    F_n [\phi_m(x),x], \qquad K^{\mu} = K^{\mu} [\phi_m(x),x]
\end{equation}
are local functionals of the fields and possibly also of the coordinates $x$. A symmetry for which neither $F_n$ nor $K^\mu$ depend explicitly on the coordinates is said to be `uniform'~\cite{WB}. This is typically the case for internal symmetries.

By Noether's theorem, the conserved current is expressed in terms of ${\cal L}$ and $K^\mu$ as
\begin{equation}
J^\mu = - K^\mu + \frac{\partial \cal L}{\partial (\partial_\mu \phi_n)} \delta \phi_n - \partial_\nu \frac{\partial \cal L}{\partial (\partial_\nu \partial_\mu \phi_n)} \delta \phi_n + \dots 
\end{equation}
Note that, although we are not considering explicitly coordinate-dependent Lagrangians, the current can inherit explicit coordinate dependence from $K^\mu$, as for instance in the case of boost invariance.
The classical analogue of \eqref{ssp} would now be an SSP  solution ${\bar \phi}_n (t)$, with
\begin{equation} \label{classic}
\dot {\bar \phi}_n (t) = c \,  F_n [\bar \phi_m(t),t] .
\end{equation}
Again, time translations and the symmetry under consideration are separately broken by the solution, but they produce the same change on the field configuration. 
A familiar example of this situation in classical physics is circular orbits of a point particle in the presence of a central force field. Such a system is invariant under rotation and angular momentum is conserved for any solution. Circular orbits are special because rotations about their axis and time evolution act on them in the same way. 

At the classical level, the existence of SSP solutions is not quite guaranteed by the symmetries of the theory, but it almost is. In the presence of a continuous symmetry, solutions can be found by inserting a symmetric ansatz into the action and extremizing within such a restricted set of symmetric  configurations \cite{coleman}. In our case the symmetries we would like to impose on the solution are the unbroken ones: spacial translations and rotations, and the unbroken linear combination of time translations and internal symmetry transformations. Typically, condition \eqref{classic} boils down to an algebraic equation allowing real solutions at least in some regions of parameter  space.


\subsection{Examples}\label{examples}

It is useful, also for later reference, to mention here some simple examples of SSP states. 

\vspace{6pt}

\noindent
{\bf Example 1.}  \emph{  A free massless scalar} with Lagrangian
${\cal L} = -\frac{1}{2} \partial_\mu \phi \partial^\mu \phi$ gives the simplest example of SSP.
the Lagrangian is invariant under the shift symmetry $\delta \phi = a$,
with constant $a$. The corresponding Noether's current is
\begin{equation}
J^\mu = \frac{\partial {\cal L}}{\partial (\partial_\mu \phi)} \delta \phi = - a\,  \partial^\mu \phi .
\end{equation}
The shift symmetry is always spontaneously broken---it is non-linearly realized on the  $\phi$ field. With time-dependence, any homogeneous configuration with $\dot \phi = {\rm const}$ is a solution, and obeys \eqref{classic}.

\vspace{6pt}

\noindent
{\bf Example 2.}
 The \emph{$U(1)$ linear sigma model} is defined by the Lagrangian
\begin{equation} \label{sigmamodel}
{\cal L} = - \sfrac12 \partial_\mu \Phi \partial^\mu \Phi^* + \sfrac{1}{2} \mu^2 \, |\Phi|^2 - \sfrac{1}{4} \lambda  \, |\Phi|^4 \; ,
\end{equation}
which is invariant under the $U(1)$ symmetry $\Phi  \to e^{i a}  \Phi$.
After the field redefinition $\Phi = \sigma e^{i \theta}$, 
the $U(1)$ symmetry acts just as a shift for the $\theta$ field: $\theta \rightarrow \theta + a$. The corresponding current is simply
\begin{equation} \label{current}
J^\mu = \frac{\partial {\cal L}}{\partial (\partial_\mu \theta)}  = -   \sigma^2 \partial^\mu\theta \; .
\end{equation}
The equations of motion admit ``rotating", SSP solutions of the form
\be
\bar \sigma(t) = v = {\rm  const} \; ,  \quad \bar \theta (t) = c t \; ,
\ee
provided we adjust the value of the radial field $\sigma$ to account for the ``centrifugal force",
\be \label{values}
 c^2 v + \mu^2 v - \lambda v^3 = 0\, .
\ee


\vspace{6pt}

\noindent
{\bf Example 3.}
 \emph{Superfluids.} A superfluid can be defined as a system carrying a conserved $U(1)$ charge, in a state that {\em (i)} spontaneously breaks the corresponding symmetry, and that {\em (ii)} has finite charge density. As we will see below, such a combination is the trademark of SSP.  At lowest order in the derivative expansion, the low-energy effective action for a (relativistic) superfluid is \cite{son}
\be
{\cal L} = P(X) \; , \qquad X =(\partial \psi)^2 \; ,
\ee
where $\psi$ is a scalar field that non-linearly realizes the $U(1)$ symmetry: $\psi \to \psi + a$. The ground state at finite charge density corresponds to the classical solution
\be \label{super}
\bar \psi(x) = \mu t \; ,
\ee
where $\mu$ is the chemical potential \cite{son}. Eq.~(\ref{super}) is clearly an SSP solution. The linear sigma model of Example 2 above can be thought of as a possible UV-completion of this low-energy effective theory, with the angular variable $\theta$ playing the role of $\psi$.

\vspace{6pt}

\noindent
{\bf Example 4.}
\emph{Dilations}. Classical scalar field theories without massive parameters are invariant under (weight 1) dilations
\begin{equation} \label{conf}
\delta \phi = (1+ x^\mu \partial_\mu) \phi \; .
\end{equation}
For definiteness, let us consider Lagrangians of the form
\begin{equation}
{\cal L} = \phi^4 f \left(\frac{X}{\phi^4}\right) ,
\end{equation}
where $X = (\partial \phi)^2$, and $f$ is an arbitrary function. Relevant examples are a $\lambda \phi^4$ theory, and the DBI Lagrangian for a probe brane in pure $AdS_5$, which have $f(x) = - x/2 - \lambda/4$ and $f(x) = \sqrt{1 + x} -1$  respectively. A higher-derivative generalization where $f$ involves second derivatives of $\phi$ as well is provided by the conformal galileon \cite{NRT, dRT}. 
Under \eqref{conf} the Lagrangian transforms like 
\begin{equation}
\delta {\cal L} =  \, \partial_\mu (x^\mu {\cal L} ) \;  
\end{equation}
which yields
\begin{equation} 
J^\mu = 2 (\phi + x^\nu \partial_\nu \phi) f' \partial^\mu \phi - x^\mu \phi^4 f 
\end{equation}
as Noether current.
Note that in this case the charge is explicitly time-dependent.

The SSP condition \eqref{classic} here reads
\begin{equation} \label{ansatzssp}
\dot \phi (t) = c(\phi + t \dot \phi) \quad \Rightarrow \quad \phi(t) = \frac{A}{1 - c t} \; ,
\end{equation}
for some constant $A$. 
The eom of a $\lambda \phi^4$ theory admit SSP solutions for negative values of $\lambda$ only, $\lambda = - 2 c^2/A^2$. Interestingly, this is the case considered in \cite{rubakov,justin}.   
For DBI instead, we find $A = c^2$.  However, this corresponds also the to speed limit in the 5D bulk~\cite{dbi}. More generally, we find that SSP solutions are a good approximation to the small speed of sound DBI regime. For the conformal galileon, SSP solutions can violate the null energy condition without instabilities \cite{NRT, genesis}.


\subsection{Plan of the paper, and summary of our results}

The plan of the paper is as follows:

In Sec.~\ref{general} we describe the properties of SSP states that follow purely from symmetry considerations, and we draw a precise connection between our SSP phenomenon and  that of standard spontaneous symmetry breaking (SSB): {\em if} the theory features SSB in the usual sense, one can construct SSP states as zero-momentum Goldstone boson coherent states. But we also argue that SSP is a more general phenomenon.

In Sec. \ref{sec.3} we extend the Goldstone theorem to our time-dependent field configurations and show that, in the case of internal symmetries, the state $|c\ket$ admits a gapless excitation. As in the Lorentz invariant case, the conserved current interpolates the Goldstone states and {\em defines} the Goldstone field operator $\pi(x)$. By studying the transformation properties of $\pi(x)$ we show that the low-energy effective field theory is always equivalent to a $P(X)$ theory expanded about a time-dependent background. Finding a massless excitation in the presence of a spontaneously broken symmetry might not come as a surprise. However, this result is nontrivial. In fact, when the symmetry under consideration is part of a larger non-abelian group, and there are other generators that are broken by the SSP state, many of the associated Goldstone excitations become massive. In a sense, the time-dependent background `higgses' some of the Goldsones. This is a very general result that we summarize and exemplify in Sec. \ref{sec.4}, while a general proof will be given elsewhere~\cite{NP}. 

There is a sense in which the  SSP state's time-dependence  is fake. We are evolving in time, but we are moving along a symmetry direction. At least for internal symmetries, this really means that perturbations of such a state feature time-independent dynamics. This result is proven in full generality in Sec.~\ref{sec.5}, by finding a convenient (time-dependent) parameterization of field space. As an illustration of our method, we apply such a parameterization first to general linearly realized symmetries, and then to $N$ scalars in the fundamental representation of $SO(N)$. Such a semiclassical analysis confirms our prediction that certain Goldstone fields acquire a gap.

\section{General properties of SSP states}\label{general}

As usual with symmetric theories, there are a number of properties that follow purely from the symmetry structure of the theory, regardless of the details of the dynamics. This is especially true for spontaneously broken symmetries, where the Goldstone theorem ensures that certain excitations are gapless and zero-energy theorems constrain their interactions at low energies. We will adapt the Goldstone theorem to our case in the next section. Here, we want to derive a number of general features for the SSP states themselves---in the absence of excitations---that follow purely from the symmetry breaking pattern.

\subsection{The $c$'s are constant}
It is easy to see that, if \eqref{ssp} holds, the parameter $c$ needs to be constant in time. By deriving \eqref{ssp} on both sides with respect to time, and with the aid of \eqref{chargecons},  we obtain
\begin{equation}
({\dot H} - {\dot c} \, Q) \, |c \ket = 0 .
\end{equation}
For a relativistic field theory in Minkowski space, the Hamiltonian is not explicitly time dependent. In this case ${\dot H} = \partial H/\partial t = 0$. It then follows that ${\dot c} = 0$. (Note that $Q |c\ket \neq 0$, because we are assuming that the symmetry is spontaneously broken in the first place.) 

In the case of eq.~(\ref{sspmany}), where there are several spontaneously broken symmetries,
it is not restrictive to assume that all the $Q_i |c\ket$'s in the sum are linearly independent vectors: If there existed a vanishing linear combination of them, it would define an unbroken symmetry. Via a change a basis we could eliminate it from our discussion and focus on the broken combinations of the original charges.
Again, we could have in principle a situation where the different $c_i$ are time dependent coefficients. To see that in fact this cannot happen, we derive \eqref{sspmany} with respect to time and obtain
\begin{equation}
 \sum_ i {\dot c}_i\, Q_i |c\ket\, =0 .
 \end{equation}
Since we have assumed that the $Q_i |c\ket$'s are all linearly independent, we need to have again ${\dot c}_i =0$. This also means that we can always find a new basis of the Lie algebra such that time evolution is parallel to just one symmetry generator $Q \propto \sum_i c_i Q_i$. This is why condition \eqref{ssp} is not restrictive, even in the presence of more than one continuous symmetry.

\subsection{Lorentz boosts are broken}
It is obvious that our SSP states break all Lorentz boosts. Indeed, they break time-translations without breaking spacial ones. This of course selects the time direction as special. More formally, given the Poincar\'e algebra, and in particular
\be
[K_i, P_j] = - i H \, \delta_{ij} \; ,
\ee
we have
\be
P_j K_i  \, | c \ket = i \delta_{ij} \, H  |c \ket \; ,
\ee
where we used the homogeneity of our states, $P_j |c \ket = 0 $. Since the r.h.s.~is nonzero, we must have
\be
K_i  | c \ket  \neq 0 \; .
\ee


\subsection{SSB + finite charge density = SSP}\label{SSB+rho=SSP}
Consider a theory whose vacuum exhibits spontaneous symmetry breaking (SSB) in the usual sense, i.e.~without breaking any of the Poincar\'e generators. There will be Goldstone bosons, one for each broken generator. We can focus on one of these Goldstones---let's call it $\pi$---and treat the corresponding generator as a $U(1)$ generator that acts on $\pi$ as a shift. We can always do this, even when the full symmetry group is non-abelian. In other words, one of the broken symmetries can always be realized as a constant shift on one of the Goldstones and as the identity on the other ones---perhaps this choice will make the action of the other symmetries more complicated, but this is not relevant for the present discussion. Now consider a state $|\psi\ket$---if it exists---that has finite charge density in such a symmetry direction:
\be \label{finitecharge}
\bra \psi | J^0 | \psi \ket  \equiv \rho_0 = {\rm const} \; ,\qquad \bra \psi | \vec J \, | \psi \ket  = 0\; .
\ee
We will actually construct such a state below. The low-energy effective Lagrangian for the Goldstones reads
\be
{\cal L}_{\rm eff} = -\sfrac12 f^2 (\partial \pi)^2 + \dots \; ,
\ee 
where the dots stand for terms that are higher order in $\pi$ (i.e., interactions) or that involve the other Goldstones, which do not transform under the symmetry we are focusing on. Therefore, at lowest order in the Goldstones, the current associated with our symmetry is simply
\be \label{j=dp}
J^\mu = f^2 \partial^\mu \pi + \dots \; .
\ee
Indeed, because of Goldstone theorem, the current has to interpolate the corresponding Goldstone:
\be
\bra 0 | J^\mu (x)| \vec p \ket = i f \, p^\mu \, e^{i p\cdot x} \; ,
\ee
which is consistent with \eqref{j=dp} (notice that the canonically normalized field is $f\cdot  \pi$). Now, for small enough charge densities we can stick with the lowest order expression \eqref{j=dp} and rewrite \eqref{finitecharge} as
\be
\bra \psi | \dot \pi (x)| \psi \ket = {\rm const}  \; ,\qquad \bra \psi | \vec \nabla \pi(x) | \psi \ket  = 0 \; ,
\ee
or more to the point:
\be \label{pi=t}
\bra \psi | \pi (x)| \psi \ket = {\rm const}  \cdot t \; .
\ee
Time translations are broken; spacial ones are not; the internal symmetry was broken to begin with; time translations shift $\pi$ by a constant, and so does the shift symmetry of course. We thus reach the conclusion that---at least for small enough densities---states of finite charge densities in a theory with standard SSB are SSP states.

Apropos of a finite charge density, it is interesting to interpret eq.~\eqref{ssp_lambda} in thermodynamical terms. According to that equation, $|c \ket$ is the ground state of $(H - c Q)$. That is, it describes the state of the system in equilibrium at zero temperature but finite chemical potential for the charge $Q$, with $\mu = c$. The eigenvalue $\lambda$ is then related to pressure $p$, since
\be
E - \mu \cdot Q = - p \cdot V \; .
\ee
With this interpretation in mind, for the superfluid example of sect.~\ref{examples} one gets
\begin{align}
H & = \int \! d^3 x \big[2 P'(X) \dot \psi^2 - P(X)\big] \; , \qquad p  = P(X)  \\
Q & = \int \! d^3 x \, 2 P'(X) \dot \psi \; , \qquad c  = \dot \psi \; .
\end{align}


\subsection{Any theory with SSB admits SSP\dots}
We now show that these finite charge density states do exist in any theory exhibiting standard SSB. We will show this by constructing the states explicitly, as suitable zero-momentum Goldstone boson coherent states. Although Goldstone bosons have the dispersion relation of a relativistic massless particle, $E(k) = |\vec k|$, for reasons that will be clear in the following we keep the discussion more general and allow generic ``massless"---in the sense that $E(0) = 0$---dispersion relations. 

The Goldstone bosons are weakly coupled at low energies. This means that as long as we stick with zero momentum particles, we can treat them as free. Notice that, however, a coherent state will involve infinitely many of them, so this conclusion may be too quick. We will discuss this subtlety below, and ignore it for the moment.
For a free field, associated with any classical solution $\bar \pi(x)$ there is a coherent state
\begin{equation} \label{coherent}
|\bar \pi(x)\rangle = {\cal N} \exp\left(\int d^3 k \ \eta_{ \vec k} a^\dagger_{\vec k}\right) |0\rangle
\end{equation}  
such that the field operator $\pi(x)$ has expectation value $\bar \pi(x)$. Here ${\cal N}$ is a normalization factor, and the $\eta_{\vec k}$'s are just the Fourier coefficients of the classical solution: 
\begin{equation} \label{expand}
\bar \pi(x) = \int d^3 k \,  \frac{\eta_{\vec k} }{\sqrt{2 E(k)}} \, e^{i (\, \vec k \cdot \vec x - E(k) t \,)} + {\rm c.c.} \; ,
\end{equation}
with $k \equiv |\vec k  |$ and $E(k)$ is the dispersion relation.
It is straightforward to check by direct inspection that $\bra \bar \pi(x)| \pi(x)| \bar \pi(x)\rangle = \bar \pi(x)$. Therefore, if we want to build a quantum state representing a given classical solution, we just have to expand such a solution as in \eqref{expand} and then plug the   $\eta_{\vec k}$ coefficients thus found into \eqref{coherent}. For a spatially homogeneous field,  $\eta_{\vec k}$ must be supported at $\vec k= 0$ only, i.e.~it must be proportional to a Dirac-delta function or to derivatives thereof. For the solution under study---that corresponding to eq.~\eqref{pi=t}---we can write the mode expansion as 
\begin{equation}
\bar \pi(x) = -\frac{i c}{4\pi} \int d^3 k \frac{\delta'(k)}{k^2 E'(k)} e^{i (\kk \cdot \x - E t)} + {\rm c.c.} = c \, t \; .
\end{equation}
The integral is straighfoward to compute by first integrating over the angular directions
\footnote{There is a divergent boundary term---proportional to a $\delta$-function evaluated at $k=0$---which disappears once we take the real part. Other divergent terms arise in the case in which $E(k)$ has vanishing first derivative at $k=0$. If $E \propto k^2$ at low momenta, these divergences do not enter the real part either. However in such a case there is a finite correction to the $c$ coefficient on the right hand side.}.
The result is valid only as long as $E(k)$ vanishes in the $\vec k \rightarrow 0$ limit. We thus have
\be
\eta_{\vec k} = -\frac{i c}{2\pi\sqrt 2 } \, \frac{E(k)}{k^{2} E'(k)} \,\delta'(k) \; .
\ee
Using this in \eqref{coherent}, we get the desired SSP  state.

As we emphasized already, for this construction to work, it is crucial that we can approximate the dynamics of the Goldstone bosons as free. Otherwise, we could not take multi-particle states as in \eqref{coherent} hoping to define this way a sensible stationary state. It is easiest to assess the importance of interactions
directly in terms of the classical field solution (interpreted as the expectation value of the quantum field operator), from the effective Lagrangian
\be
{\cal L} \sim f^2 (\partial \pi)^2 + \alpha (\partial \pi)^4 + \dots \; ,
\ee
where $\alpha$ is a coupling constant the depends on the UV-completion.
If there are several broken symmetry generators, there are in general interaction terms with two derivatives coupling $\pi$ to the other Goldstones, but since these are not excited in the state under study, such interactions will give subleading effects w.r.t.~to those captured by the  interaction explicitly displayed. For a state with constant $\dot \pi$, the interaction term becomes important when
\be
\dot \pi^2 \sim f^2/\alpha \; .
\ee
This does not necessarily mean that the theory becomes (quantum mechanically) strongly coupled at this point, or that the low-energy effective field theory has to break down there \cite{CN}
\footnote{See also a related discussion in \cite{NR}.}---it simply means that our simple minded construction of the SSP state does not work for such high $\dot \pi$'s.
On the other hand, for substantially smaller $\dot \pi$'s the dynamics of $\pi$ can be safely treated as free, and our coeherent state \eqref{coherent} is a perfectly sensible SSP state.

Finally, as we show in Sec.~3,  in the vicinity of an SSP state the spectrum contains a weakly coupled, gapless excitation---or ``running Goldstone"---characterized, in general, by a Lorentz breaking dispersion relation. Typically, $E \simeq c_\pi |\vec k|$ at low momenta. We can use a coherent state of these running Goldstones to go, perturbatively, from an SSP state with a given $c$ to another one with a slightly different $c$. Analogously to the above formulas, we can thus write 
\begin{equation} 
|c+\Delta c \rangle = {\cal N} \exp\left(\int d^3 k \ \eta_{ \vec k} a^\dagger_{\vec k}\right) |c\rangle,
\end{equation}  
with
\be
\eta_{\vec k} = -\frac{i \Delta c}{2\pi\sqrt 2 } \, \frac{E(k)}{k^{2}  E'(k)}\,  \delta'(k)  \; .
\ee

\subsection{\dots but not vice versa}

One can have SSP in the absence of standard  SSB. What we mean is that there are cases where the Poincar\'e invariant  state of the theory is symmetric, and yet we can consider time-dependent symmetry probing states. Take for instance the $U(1)$ linear sigma model discussed above (Example 2) in the unbroken phase, i.e.~with a  positive $m^2$ term in the potential. The vacuum preserves the $U(1)$ symmetry, but we are still allowed to have rotating solutions of the type we discussed.

Perhaps superfluids offer the most relevant physical example here. The spontaneous breaking of the $U(1)$ charge is triggered by having a finite charge density: He$_4$ atoms undergo Bose-Einstein condensation at low temperatures only if there are He$_4$ atoms around. In the absence of a  non-zero density, the state of the system is just the usual vacuum, Poincar\'e- as well as $U(1)$-invariant.
There are no standard SSB states---only SSP ones.

A more exotic possibility---which is easy to cook up classically or as a consistent low energy effective field theory, but which is not obviously realized in UV-complete theories, or in nature---is that there is no Poincar\'e invariant vacuum at all.
It is not difficult to conceive a somewhat ad-hoc $U(1)$ sigma model similar to that of Example 2 above where, however, the potential is unbounded from below and goes to $-\infty$ for $\Phi \to 0$. Such a theory does not have a vacuum in the usual sense but there will exist, classically, a charged sector where the ``centrifugal force'' will prevent the system from falling into the ``hole" at the origin. The effective field theory for perturbations about such SSP states is perfectly well behaved at low energies. A similar example is that of ghost condensation \cite{ghost, ghost2}. There, however, since the ghost condensate point has somewhat degenerate properties, the relevant SSP state has actually {\em zero} charge density. This does not contradict our analysis of sect.~\ref{SSB+rho=SSP}: there we were starting from an SSB vacuum and we were considering, in perturbation theory, charged states close to it. The ghost condensate point does not belong to this class.


\section{The Running Goldstone} \label{sec.3}

The most general proof of the Goldstone theorem~\cite{Goldstone} does not make use of the effective potential, and also applies to cases where, like for the pions, the Goldstone excitations are not directly associated with the fields appearing in the fundamental  Lagrangian. Along the same lines, we now show how  to adapt the Goldstone theorem to our SSP states. We will see that 
\emph{if there is a state $|c\rangle$ obeying \eqref{ssp} and the broken charge is not explicitly time-dependent, then $|c \ket$ admits gapless excitations in the zero momentum limit.} Like for the standard Goldstone theorem, such excitations are interpolated by the current, in the sense that single particle states of (low) momentum $\vec p$ can be created by applying $J^\mu$ to $| c \ket$:
\be
\langle c| J^\mu (x) | \vec{p} \rangle \propto e^{i( \vec{p} \cdot \vec x - E_p \, t)} .
\ee

Notice that since Lorentz boosts are spontaneously broken by $| c \ket$, in general  such excitations will have a non-relativistic dispersion law, $E_p \neq |\vec p|$. Also, like for the standard {\em non-relativistic} Goldstone theorem, at finite momentum they need not be asymptotic states of the theory, i.e.~they may be unstable against decaying into lower energy modes. This is the case for instance for the phonons of superfluid helium, which undergo the decay $\pi \to \pi \pi$ with a rate $\Gamma \propto E^5$ (see e.g.~\cite{landau}). 

We will then derive how the broken symmetries are non-linearly realized on the Goldstone field operator. This will allow us to discuss the systematics of the Goldstone low-energy effective field theory.


\subsection{The Goldstone theorem}\label{goldstone theorem}

Consider the matrix element $\langle c |[Q(t),A(0)] | c \rangle$ where $A(0)$ is some local field operator evaluated at $x=0$.  $Q$  in principle depends on $t$ because we are in the Heisenberg picture, but we will assume that it has no {\em explicit} time dependence.  We want to show that such a matrix element is time independent. By current conservation we have 
\begin{eqnarray}
0 & = & \int d^3 x \langle c| [\partial_\mu J^\mu (\vec{x},t) , A(0)] |c\rangle \nonumber \\
& = & \int d^3 x \langle c| [\dot J^0 (\vec{x},t) , A(0)] |c\rangle + \int d^3 x \langle c| [\partial_i J^i (\vec{x},t) , A(0)] |c\rangle . 
\end{eqnarray}
The last term is a boundary term that only receives contributions from spatial infinity. Because the commutator of local operators is null outside the light cone, such a term is guaranteed to vanish. Incidentally, the Goldstone theorem is known to fail in some non-relativistic cases precisely because of the lack of a light-cone structure and because of the presence of long-range (instantaneous) interactions (see e.g.~\cite{kibble, NC, lange, brauner}). Here, however, Lorentz invariance is broken by the state $|c\ket$, not by the dynamics, and the general properties of local relativistic operators apply as usual
\footnote{It is tempting to speculate that some general results of non-relativistic many-body theories could be strengthen by exploiting analogous arguments. After all, 
Lorentz invariance is broken always \emph{spontaneously} in the real world, while the underlying fundamental interactions are fully relativistic. We leave explorations along this direction for future work \cite{NP}.}.

We have shown that $\langle c |[Q(t),A(0)] | c \rangle$ is a constant. As we are assuming that $|c \ket$ breaks the symmetry generated by $Q$, such a matrix element must also be non-zero for some local (hermitian) operator $A$. In other words, 
\begin{align}
0 \neq {\rm const.} &= \int d^3 x \langle c| J^0 (\vec{x},t) A(0) |c\rangle  - {\rm c.c.} \nonumber \\
&= \int d^3 x \langle c| e^{i (P\cdot \vec{x} + H t)} \, J^0 (0) \, e^{-i (P\cdot \vec{x} + H t)} A(0) |c\rangle  - {\rm c.c.}
 \label{step}
\end{align} 
Note that the last line has been obtained by assuming that $J^\mu$ is not explicitly coordinate dependent.
 This assumption---as well as that concerning $Q$'s  explicit time-dependence---will be generically violated for spacetime symmetries that do not commute with $P^\mu$, like e.g.~the dilation invariance of Example 4 above. 

If $|c \ket$ obeys \eqref{ssp} and does not break spacial translations, the matrix element inside the integral in~\eqref{step} can be rewritten as
\begin{eqnarray}
& & \langle c| e^{i c\, Q t} \, J^0 (0) \, e^{-i (P\cdot \vec{x} + H t)} A(0) |c\rangle  \nonumber \\
& = & \sum_n \int d^3 p  \langle c| J^0 (0) e^{- i (H - c\, Q) t} e^{-i \vec{p} \cdot \vec{x}} \, |n,\vec{p} \, \rangle \langle n,\vec{p} \, | \, A(0) |c\rangle \; , \label{commute}
\end{eqnarray}
where we have summed over intermediate momentum eigenstates, ${\vec P} |n,\vec{p} \, \rangle = \vec{p} \, |n,\vec{p} \, \rangle$---$n$ labels different  sectors (e.g., multi-particle states) within a given  eigen-space of $\vec P$---and we have used that $Q$ commutes with $J^0$. Integration over $\vec x$ thus gives
\begin{equation}
{\rm const} = (2 \pi)^3 \sum_n \int d^3 p \, \delta^3(\vec{p}) \langle c| J^0 (0) e^{- i (H - c\, Q) t} \, |n,\vec{p}\rangle \langle n,\vec{p}| \, A(0) |c\rangle - {\rm c.c.} \label{commute2}
\end{equation}
Since $\vec P$ commutes with $H$ and $Q$, it is possible to choose the $|n,\vec{p}\rangle$ states in such a way that they are also eigenstates of $H-cQ$. Note that for $|n,\vec{p}\rangle = |c\rangle$ the above expression just gives zero, because $J^0$ and $A$ are hermitian operators. In order for it to be time independent \emph{and} different from zero, there must exist a state other than $|c\rangle$ that, in the limit of zero momentum, obeys the same equation \eqref{ssp}. In other words, by introducing the effective Hamiltonian 
\begin{equation} \label{tildeH}
\tilde{H} = H - c \, Q ,
\end{equation} 
which generates the unbroken linear combination of time translations and internal symmetry, 
there must be eigenstates of ${\vec P}$ in the theory, $|\pi(\vec p)\ket$, such that 
\begin{equation}
\tilde{H} |\pi(\vec p)\ket \rightarrow 0 \qquad {\rm as} \qquad \vec{p}\rightarrow 0.
\end{equation}
Moreover, the matrix elements  $\langle c| J^0 (0) \, |\pi(\vec p)\ket$ and $\bra \pi(\vec p)| \, A(0) |c\rangle$ should be different from zero. 
These states are our Goldstone bosons. 

\subsection{The Goldstone field} \label{sec4.2}

As in the Lorentz invariant case, the current operator interpolates the state $ |\pi(\vec p)\ket$. Given the residual symmetries, we expect
\footnote{We are switching to the so-called relativistic normalization for single particle states,
\be
\bra \pi(\vec q) | \pi(\vec p) \ket = 2E_{p} \, (2 \pi)^3 \, \delta^3(\vec q - \vec p) \; , 
\ee
which is {\em not} the same as that implicitly assumed in eqs.~\eqref{commute}, \eqref{commute2}:
\be
\bra n, \vec q \, | n', \vec p \,  \ket = \delta_{n n'} \, \delta^3(\vec q - \vec p) \; .
\ee
}
\begin{equation} \label{matrix}
\langle c| J^\mu (x) |\pi(\vec{p})\rangle = i \, e^{i(\vec{p}\cdot \vec{x} - E_{p} t) } f(p^2) k^\mu,
\end{equation}
where $E_{p}$ is the  eigenvalue of $\tilde H$, and
\be
k^\mu \equiv \big( E_{p}, \alpha(p^2) \, \vec p \,  \big) \; .
\ee 
Notice that, since  Lorentz boosts are spontaneously broken, $k^\mu$ needs not  be the same as $p^\mu \equiv (E_{p}, \vec{p} \,)$. However, since spacial rotations are unbroken, the spacial part of $k^\mu$ has to be aligned with $\vec p$. 
Current conservation implies
\begin{equation}
f(p^2) \big( E^2_{p}  - \alpha(p^2) \, p^{2} \big)= 0\, .
\end{equation}
Unless $f(p^2)$ vanishes, which would violate last subsection's result, the combination in parentheses has to vanish. This gives us the Goldstone dispersion law. We can assume that the dispersion law starts linear at low momenta,
\be
E_{p} \simeq c_\pi p \; ,
\ee
in which case $\alpha (p^2) \simeq c_\pi^2 $, but our results below will also apply to degenerate cases---like the ghost condensate one---in which the low-energy dispersion relation is of higher order in momentum.
At low momenta we can Taylor-expand $f$ too,
\be
f(p^2) = v + {\cal O}(p^2) \; .
\ee
We can interpret $v$ as a symmetry breaking scale. At lowest order in $p$ we thus get
\be  \label{deformed}
\langle c| J^\mu (x) |\pi(\vec{p})\rangle \simeq i v \,  e^{i(\vec{p}\cdot \vec{x} - E_{p} t) }  \big( E_{p}, c_\pi^2 \vec p \,  \big) 
\ee

Now we want to use $J^\mu$ to define the Goldstone field operator, $\pi(x)$. In perturbation theory about our SSP state, it is natural to assume an expansion of $J^\mu(x)$ in powers of the fields as follows:
\begin{eqnarray} \nonumber
J^\mu (x)&=& J^\mu_{(0)} +J^\mu_{(1)}(x) +J^\mu_{(2)}(x) + \dots \\[2mm]
&\equiv& J^\mu_{(0)} + v \, D^\mu \pi(x) + {\cal O}({\rm fields}^2) \; , \label{pi}
\end{eqnarray}
where $D^\mu$ is a four-vector differential operator which we will discuss shortly.
The presence of a constant, zeroth-order piece, $J^\mu_{(0)}$, is required by our SSP state's having---in general---finite average charge density.
Then, whatever is responsible to interpolate between $|c\ket$ and $|\pi(\vec{p})\rangle$ in \eqref{matrix}, will be linear in the field operator that we want to define. Higher order terms do not contribute to \eqref{matrix}. By demanding canonical normalization for $\pi(x)$, 
\begin{equation} \label{pi2}
\langle c| \pi (x) |\pi(\vec{p})\rangle =  e^{i(\vec{p}\cdot \vec{x} - E_{\vec p} t) } ,
\end{equation}
we get that at lowest order in derivatives the $D^\mu$ operator must reduce to
\be
D_\mu = \big( \di_t, c_\pi^2 \vec \nabla \big) + {\cal O }(\di^2) \; .
\ee
Notice that the $\pi(x)$ field is defined by \eqref{pi} and \eqref{pi2} only up to non-linear field redefinitions of the form 
\begin{equation} \label{redef}
\pi \rightarrow \pi'=\pi + {\cal O}(\pi^2) \; , 
\end{equation}
where the non-linear piece can in principle include derivatives. 

\subsubsection{The $U(1)$ sigma model}

Before proceeding with our general analysis, it is worth pausing to see how what we derived so far applies in practice to a concrete example. For this purpose  we go back to Example 2 of Sec.~\ref{examples}---the $U(1)$ linear $\sigma$-model.  We found SSP solutions of the form $\bar \theta(t) = c t$, $\bar \sigma = v = \frac1{\sqrt{\lambda}}\sqrt{\mu^2 + c^2}$ for the angular and radial fields respectively. By expanding in small fluctuations around such a solution, 
\begin{eqnarray}
\sigma(x) &=& v + s(x) \\
\theta(x) &=& c \, t + \pi(x)
\end{eqnarray}
we can write the current in terms of the fluctuations, 
\begin{equation}\label{ccurrent}
J^\mu = - c \, \delta^\mu_0  v^2  - 2 c \, \delta^\mu_0 v \, s - v^2 \partial^\mu \, \pi.
\end{equation}
We note the same structure as in \eqref{pi} with a constant term and a piece linear in the fields that contains both $\pi$ and the heavy field $s$. The Goldstone field, as we have defined it above, is thus a combination of the two. 
%

The Lagrangian for the fluctuations reads
\begin{eqnarray} \label{perturbations}
{\cal L} &=& \sfrac{1}{2}(- (\partial s)^2 - M^2 s^2)  - \lambda v \, s^3 - \sfrac14 {\lambda} s^4 \nonumber \\
&& -\sfrac12 {v^2} (\partial \pi)^2  + 2 c v \, s {\dot \pi} - v \, s \, (\partial \pi)^2   \nonumber \\
&& + c\,  s^2 {\dot \pi} - \sfrac{1}{2} s^2 (\partial \pi)^2  \, ,
\end{eqnarray}
where we have defined $M^2 = 2 (\mu^2 + c^2)$. At energies well below $M$ it is useful to integrate out the $s$ field. In first approximation, we do this at tree level by considering, among the interactions between $\pi$ and $s$, only those that are linear in $s$. Effectively, this amount to plug into the Lagrangian~\eqref{perturbations} and into the current~\eqref{ccurrent} the solutions of the equations of motions for $s$, $s = 2 c v \, \dot \pi /M^2 + {\cal O}(\partial^2, \pi^2)$. In terms of a newly defined canonical field $\pi_c$ we get 
\begin{equation}
{\cal L}_{\rm eff} = \sfrac{1}{2} \left(\dot \pi_c^2 - c_\pi^2 \nabla \pi_c^2 \right) + {\cal O}(\partial^4, \pi^3) \; .
\end{equation}
where $c_\pi^2 = 1/( 1 + 4 c^2/M^2)$ and $\pi_c = v \pi/c_\pi$. In conclusion, as predicted, there is a massless excitation with a non-relativistic dispersion law.

\subsection{Transformation properties of $\pi$}

We now want to understand how the $\pi$ field transforms under the symmetries of the theory.  To begin with, we notice that the time evolution of the operator $\pi$ is governed by the effective Hamiltionian \eqref{tildeH} rather than by the original one, in the sense that
\begin{equation} \label{evoforpi}
\frac{d \pi(x)}{d t} = i[{\tilde H},\pi(x)] .
\end{equation}
The reason is that $\tilde H$ generates the unbroken time-translations, which are actually a linear combination of the original time translations and of the internal symmetry generated by $Q$. Perturbations about $| c \ket$ can be classified in terms of eigenstates of $\tilde H$ (this is the case for our Goldstone particles). The field operators that create and annihilate these states then evolve---in Heisenberg picture---according to eq.~\eqref{evoforpi}. In Examples 1, 2, and 3 of sect.~\ref{examples}, apart from obvious adjustments in the notation, one is expanding about a field configuration $\bra \phi \ket = \alpha \cdot t$. One then defines the fluctuation field as
\be \label{piexamples}
\pi(x) \equiv \phi(x) - \bra \phi \ket = \phi(x) - \alpha \cdot t \; ,
\ee
which---like all operators---evolves in time according to
\be
\frac{d \pi(x)}{d t} = i[H,\pi(x)]  + \frac{\di}{\di t} \pi(x) \; .
\ee
The explicit time-dependence in \eqref{piexamples} precisely combines with the $[H,\pi]$ commutator to yield \eqref{evoforpi}.

Then, we consider the spontaneously broken internal symmetry generated by $Q$. By definition, it should act on $\pi$ as
\begin{eqnarray}
\delta_Q \pi(x) &=& i [Q,\pi(x)] \\
&=& i \! \int \! d^3 y \, [J^0(\vec y, t),\pi(\vec x, t)] 
\end{eqnarray}
In perturbation theory, we can expand the current as in \eqref{pi}. Moreover, if $\pi$ is the canonically normalized field---in the sense of \eqref{pi2}---its conjugate momentum is $\Pi = \dot \pi + {\cal O}(\pi^2)$, as obvious from the effective Lagrangian
\be
{\cal L}_{\rm eff} = \sfrac12 \dot \pi^2 - \sfrac12 c_\pi^2 (\vec \nabla \pi)^2 + {\cal O}(\pi^3) \; . 
\ee 
We thus get
\begin{eqnarray}
\delta _Q \pi(x) &=& i v \int d^3 y [\dot \pi(\vec y, t) + {\cal O}(\pi^2),\pi(\vec x, t)] \\
&=& v + G[\pi] \; . \label{junk}
\end{eqnarray}
where $G[\pi]$ is a local functional of the $\pi$ field that starts linear in $\pi$.
Notice that such a transformation law is sensitive to field redefinitions of $\pi$ of the form \eqref{redef}. We can actually exploit such an ambiguity to set $G[\pi] \to 0$, in which case the (newly defined) $\pi$ transforms under the broken symmetry by a constant shift:
\be \label{transformQ}
\delta _Q \pi(x)  = v \; .
\ee
That this is possible is shown in the Appendix.

Next, we can combine this internal symmetry with \eqref{evoforpi} and derive the transformation of $\pi$ under the original time-translations, which are generated by $H$:
\be \label{deltaHpi}
\delta_H\pi(x) = i[H, \pi(x)] = i[\tilde H, \pi(x)] + c \cdot i[Q, \pi(x)] = \dot \pi(x) + c \cdot v \; .
\ee
On the other hand, spacial translations are not spontaneously broken, and are therefore linearly realized on $\pi$ as usual:
\be
\delta_{P^i} \pi(x) \equiv i[P^i, \pi] = \di^i \pi
\ee

Finally, combining spacial and time-translations we can determine the $\pi$ transformation law for Lorentz boosts. Indeed for any Poincar\'e invariant theory the boost generators can be expressed as
\be
K^i = t \, P^i - \int \! d^3 y \, y^i T^{00}(\vec y, t) \; ,
\ee
where $T^{\mu\nu}$ is the stress-energy tensor operator. Since $H = \int \! d^3 y \, T^{00}(\vec y, t)$, we must have
\be \label{T with pi}
i [T^{00}(\vec y, t), \pi(\vec x, t)] = \delta^3(\vec x- \vec y) \, \delta_H \pi(x) \; .
\ee
In principle there could be additional contact terms proportional to derivatives of the delta function, which would integrate to zero in $[H, \pi]$ but which would nonetheless modify the above local commutation relation. In fact such terms are not there, as we show in the Appendix. We thus get
\be
\delta_{K^i} \pi(x) \equiv   i[K^i, \pi(x)] = t \, \delta_{P^i} \pi(x) - x^i \, \delta_H \pi(x)
\ee

In conclusion the infinitesimal transformation laws of $\pi$ under the spontaneously broken symmetries are
\begin{align}
\pi & \to \pi + \epsilon \, v &  \mbox{(internal charge)}   \\
\pi & \to \pi + \epsilon \,  \dot \pi + \epsilon \, c v  &\mbox{(time translations)} \\
\pi & \to \pi + \vec \xi \cdot \big( t \, \vec \nabla \pi - \vec x \, \dot \pi \big) -   c v  \big(\vec \xi \cdot \vec x \big) & \mbox{(boosts)} 
\end{align}
whereas we have the standard linear transformations under the unbroken symmetries---spacial translations and rotations, as well as the unbroken $\tilde H$ Hamiltonian.

\subsection{The low-energy effective theory}

The non-linear transformation properties of the Goldstone field suggest a very efficient way of constructing the low-energy effective field theory for the Goldstone excitations of an SSP state. We just have to define a new field $\phi(x)$ as
\be
\phi(x) \equiv c  v \, t + \pi(x) \; .
\ee
Such a field realizes linearly---as an ordinary scalar---{\em all} spacetime symmetries, and shifts under the spontaneously broken internal symmetry:
\begin{align}
\phi & \to \phi + \epsilon \, v &  \mbox{(internal charge)}   \\
\phi & \to \phi + \epsilon \,  \dot \phi   &\mbox{(time translations)} \\
\phi & \to \phi + \vec \xi \cdot \big( t \, \vec \nabla \phi - \vec x \, \dot \phi \big)  & \mbox{(boosts)} 
\end{align}
Its low-energy effective Lagrangian therefore is
\be
{\cal L}_{\rm eff} = P(X) + \mbox{higher derivatives} \; ,\qquad X \equiv (\di_\mu \phi \, \di^\mu \phi) \; ,
\ee
where $P$ is a generic function, and the higher derivative terms are constrained just by shift invariance---i.e., each $\phi$ should carry at least one derivative---and by Lorentz invariance---implemented as usual.

Notice that introducing $\phi$ is purely a matter of convenience, and no obvious physical significance should be attached to it. For instance, $\phi$ should not be interpreted as a fundamental field that takes  a time-dependent vev in the SSP state, although in some cases it may be just that. Likewise, the effective theory is not supposed to necessarily make sense---or to be trusted---about $\phi = 0$. In fact, our derivation only shows that ${\cal L}_{\rm eff}$  is the correct low-energy description of the system for perturbation theory about the SSP state with $\bra \phi \ket = cv \, t$: it is the most general local theory involving the correct (perturbative) degrees of freedom and  obeying the correct symmetries.

\section{More broken generators: the higgsed  Goldstones} \label{sec.4}

So far we have focused on the properties of the excitation that perturbs the system along the SSP direction---the running Goldstone. We now consider more systematically the case in which our charge $Q$ is just one of the generators of a non-abelian group $G$. In general our SSP state $|c\ket$ will break other generators of $G$ as well. According to standard lore, excitations that perturb the system along these other symmetry directions---let's call them the transverse Goldstones---are also gapless. What we now show is that, in  fact, they are not. More precisely, all transverse Goldstones associated with broken generators that do not commute with $Q$ are gapped. On the other hand, all transverse Goldstones associated with broken generators that {\em do} commute with $Q$ are gapless. For brevity, let us call these two classes of generators `NC' and `C', short for `non-commuting' and `commuting'. Of all the generators of $G$, we will only consider the spontaneously broken ones.

To begin with, let us see where the Goldstone theorem fails for the NC generators. Consider the proof in sect.~\ref{goldstone theorem}, adapted to a generator $Q_a = \int d^3 x J^0_a$ other than $Q$. Eq.~\eqref{step} will now read
\begin{eqnarray}
0 \neq {\rm const} &=& \int d^3 x \langle c| J_a^0 (\vec{x},t) A(0) |c\rangle  - {\rm c.c.} \nonumber \\
&=& \int d^3 x \langle c| e^{i (P\cdot \vec{x} + H t)} \, J_a^0 (0) \, e^{-i (P\cdot \vec{x} + H t)} A(0) |c\rangle  - {\rm c.c.}
 \label{step2}
\end{eqnarray} 
The leftmost exponential can still be converted to  $e^{i c\, Q t}$---$|c \ket$ is annihilated by $\vec P$, and acting on it with $H$ is equivalent to acting with $c Q$. However now if $Q_a$ and $Q$ do not commute with each other, in order to move the exponential to the right, past $J^0_a$, we have to pay a commutator. 
This impairs all the subsequent steps in our proof, and therefore its conclusion. Of course this complication is not there for C generators, which then have gapless Goldstone excitations associated with them.

Interestingly, for the NC generators, if  we apply the same logic as in sect.~\ref{goldstone theorem}, we can derive the {\em gap} for the excitations interpolated by the associated currents.  We do this in broad generality in a forthcoming publication \cite{NP}. Here, we just quote the general result: {\em the currents associated with the NC generators interpolate excitations that are gapped in the zero-momentum limit, with gap given by $|c \, q_a|$, where $q_a$ is the generator's charge under $Q$-transformations.}




To get a more intuitive understanding of where the gap is coming from, we will now check these statements in a simple example; then, in sect.~\ref{sec.5}, we will carry out a systematic semiclassical analysis of SSP at the level of the Lagrangian for  non-abelian symmetry groups.

The alert reader may have detected a contradiction between our claim above and the Nielsen-Chadha theorem \cite{NC},
whereby for non-relativistic theories with spontaneously broken internal symmetries, there are as many {\em gapless} Goldstone bosons as the number of broken generators, provided one counts the excitations with a quadratic dispersion relation ($E \propto p^2$) twice.
We address this contradiction in ref.~\cite{NP}. For the time being, suffice it to say that our examples below precisely match our claim.

\subsection{Example: SSP in the chiral Lagrangian}

As we discussed in sect.~\ref{general}, for small enough $c$'s we can construct SSP states by exciting weak time-dependent Goldstone fields in a theory that exhibits standard SSB.  In the case of a spontaneously broken non-abelian group, the Goldstone bosons have non-trivial interactions than involve fewer derivatives than fields, e.g.~$\pi^2 (\di \pi)^2$. It is then clear that if we consider a time-dependent background field and we expand the action at quadratic order in fluctuations, we can get mass terms for these fluctuations whenever the derivatives in such interaction terms hit the background field. This is, in essence, the origin of the gap, at least in the cases in which the SSP state can be constructed perturbatively starting from an ordinary SSB vacuum.

For definiteness, consider  the $SU(2)$ chiral Lagrangian, describing the low-energy dynamics of a theory spontaneously breaking $G = SU(2) \times SU(2) $ down to the diagonal $SU(2)$ subgroup. Such a theory is conveniently parameterized in terms of the unitary matrix-valued field
\be
\Sigma (x) = e^{i \,  \sigma_a \pi_a(x)}  \;, 
\ee
where the $\sigma_a$'s are the Pauli matrices, which are hermitian, and the $\pi_a$'s are the Goldstone fields---the `pions'. The low-energy effective Lagrangian is
\be
{\cal L} = - \sfrac14 f^2 \, {\rm tr} \big[ \di_\mu \Sigma^\dagger \di^\mu \Sigma \big ] + \dots \; ,
\ee
where the dots stand for an infinite series of higher derivative terms, which include higher powers of $\di \Sigma$ as well as more derivatives acting on each $\Sigma$. 

Consider now an SSP configuration evolving along $\sigma_1$; it corresponds to a time dependent $\pi_1$ field, and vanishing $\pi_2, \pi_3$ fields. We can plug such an ansatz directly into the action  and extremize over $\pi_1(t)$ to find the actual solution: the reason is that such an ansatz preserves one of the symmetries of the action---a linear combination of $\sigma_1$-transformations and of time-translations---and is thus automatically an extremum for variations that do not respect such a symmetry (cf.~sect.~\ref{intro}). Notice that if only $\pi_1$ is excited, then $\Sigma = \exp(i \,  \sigma_1 \pi_1)$ `abelianizes', in the sense that it behaves as a phase in the Lagrangian: it commutes with $\Sigma^\dagger = \exp(-i \,  \sigma_1 \pi_1)$, with $\di_\mu  \Sigma = \di_\mu \pi_1 \, \sigma_1 \Sigma$, and so on.  The reason is that, as a matrix, it is just a function of $\sigma_1$, and so are $\Sigma^\dagger$,  $\di_\mu  \Sigma$, and so on. For our ansatz then, the Lagrangian reduces to that of a derivatively coupled scalar field,
\be
{\cal L} \to {\cal L}(\di_\mu \pi_1, \di_\mu \di_\nu \pi_1, \dots) \; ,
\ee
which always admits homogeneous time-dependent solutions with constant velocity:
\be \label{SSP pi_1}
\pi_1 = \alpha t \; , \qquad \alpha = {\rm const} \; .
\ee
We now consider small fluctuations about this SSP solution, and we are interested in the quadratic Lagrangian for such fluctuations. For small $\alpha$---i.e.~for weak background fields---we can expand the Lagrangian above in powers of the fields, and then decompose each field into background plus fluctuations. After straightforward algebra\footnote{One should use the following trace identities for the Pauli matrices:
\be
{\rm tr} [\sigma_a \sigma_b] = 2  \, \delta_{ab} \; , \quad 
{\rm tr} [\sigma_a \sigma_b \sigma_c] = 2 i \, \epsilon_{abc} \; , \quad
{\rm tr} [\sigma_a \sigma_b \sigma_c \sigma_d] = 2 (\delta_{ab} \delta_{cd } - \delta_{ac} \delta_{bd} + \delta_{ad} \delta_{bc}) \; .
\ee
}, the first step yields
\be
{\cal L} \to -f^2 \big[ \sfrac12 (\di_\mu \vec \pi)^2 + \sfrac1{6} \di_\mu\pi_a \di^\mu \pi_b \, (\pi_a \pi_b - \delta_{ab} \, \vec \pi \, ^2) +\dots \big] \; ,
\ee
where the dots stand for terms involving more than four fields or more than two derivatives, or both.
Once we replace $\pi_1$ with its background value plus fluctuations, $\pi_1 \to \alpha t + \pi_1$, and we keep up to quadratic order in fluctuations, we get
\be
{\cal L} \to  -f^2 \big[ \sfrac12 (\di_\mu \pi_1)^2 
+ \sfrac12 (\di_\mu \vec \pi_T) ^2 + \sfrac1{6} \big( \alpha^2 \, \vec \pi_T^2 - 2 \alpha^2 t \, \vec \pi_T \cdot \dot{\vec \pi}_T 
- \alpha^2 t^2 (\di_\mu \vec \pi_T)^2 \big) + \dots \big] \; ,
\ee
where the $SU(2)$ vector $\vec \pi_T$ denotes the `transverse' fields, $\vec \pi_T = (0, \pi_2, \pi_3)$.
We see that the free dynamics of $\pi_1$  are unaltered by the background: $\pi_1$ is still a massless excitation, and one should go to higher orders in the derivative expansion to see modifications to $\pi_1$'s propagation speed.
On the other hand, the dynamics of the transverse pions are affected drastically by the background, already at the order we are considering: we have a mass term for $\vec \pi_T$, a one-derivative term  that  can be rewritten as a mass term by integrating by parts, and a time-dependent modification to the kinetic term. At lowest in order in $\alpha^2$, the last correction can be reabsorbed by the time-dependent field redefinition
\be
\vec \pi_T \to \big( 1 + \sfrac1{12} \alpha^2 t^2 \big) \, \vec \pi_T  \; ,
\ee
which creates at the same order in $\alpha^2$ a further one-derivative correction, $-\sfrac1{3} \alpha^2 t \, \vec \pi_T \cdot \dot{\vec \pi}_T$.
Upon integrating by parts we are left with
\be
{\cal L} \to -f^2 \big[ \sfrac12 (\di_\mu \pi_1)^2  + \sfrac12 (\di_\mu \vec \pi_T) ^2 + \sfrac1{2} \alpha^2 \, \vec \pi_T^2 + \dots \big] \; .
\ee
That is, in the presence of a weak SSP background \eqref{SSP pi_1} the transverse pions get a mass
\be
m^2_T =  \alpha^2 \; .
\ee
This matches precisely the energy gap predicted above.

To conclude, notice also that with our original parameterization of the Goldstone fields, the Lagrangian for small perturbation about the SSP configuration depends explicitly on time. Yet as we know there is an unbroken linear combination of the original Hamiltonian and of $Q_1$, which can serve as the generator of unbroken time translations. It is natural to expect that there is a more clever parameterization of the fields that makes this explicit, that is to say, that makes the Lagrangian for perturbations explicitly time-independent. We checked this above, at quadratic order in perturbations and at lowest order in the background field. In the next section we are going to present a generic, non-linear construction of these optimal field variables without committing to any specific Lagrangians.

\section{Semiclassical analysis} \label{sec.5}

The Goldstone theorem that we have extended to SSP states is a statement about the low-energy spectrum of the charged sector of the theory. We have shown that such a sector contains (at least) one massless excitation. More properties can be explored semi-classically, by expanding the classical action around an SSP solution.

\subsection{The Lagrangian for perturbations is time-independent} \label{sec.5.1}

Consider a relativistic field theory for $N$ scalar fields $\phi_n$, obeying some global symmetries.
Fluctuations of a time-dependent solution $\bar \phi_n(t)$ are, in general, governed by an explicitly time-dependent Lagrangian: 
\begin{equation}
{\cal L}_{\rm fluc}[\varphi_n(x),t] \equiv {\cal L}[\bar \phi_n(t) + \varphi_n(x)] \; ,
\end{equation}
where ${\cal L}[\phi_n]$ is the original Lagrangian, and $\varphi_n(x) = \phi_n(x) - \bar \phi_n(t)$ are the fluctuations. 
More generally, any time-dependent field redefinition 
\begin{equation}
\phi_n = \phi_n(\varphi_m,t)
\end{equation}
will induce a `spurious' explicit time-dependence in the Lagrangian for the new fields $\varphi_n$:
\begin{equation}
\frac{\partial {\cal L}_{\rm fluc}[ \varphi_m ,t]}{\partial t} = \frac{\delta {\cal L}}{\delta \phi_n} * \frac{\partial \phi_n(\varphi_m,t)}{\partial t},
\end{equation}
where summation over repeated indices is understood, and the `star' denotes a convolution.

However, we might expect fluctuations of an SSP solution to be governed by a time-independent Lagrangian. After all, we are moving along a symmetry direction, and the action expanded at different times should thus be the same. Equivalently, there is an unbroken linear combination of time-translations and of global symmetries that can serve---in principle---as a new time-translational symmetry. In the following we show that, if the symmetry is uniform, this is indeed the case. We will find a  parameterization of field space that makes the Lagrangian for the fluctuations manifestly time-independent to all orders. 

Consider a uniform (i.e., coordinate independent) symmetry transformation $\delta \phi_n(x) = F_n[\phi_n(x)]$. By definition, 
\begin{equation} \label{symmetry!}
\frac{\delta {\cal L}} {\delta \phi_n} * F_n[\phi_m] = 0 .
\end{equation}
Consider now an SSP solution $\bar \phi_n(t)$ for this symmetry, and a neighborhood thereof in field space. Such a neighborhood is, topologically, a tube. We want to put coordinates in it by using the set of trajectories that are generated by the symmetry. Every such trajectory is a one-parameter curve $\Phi_n(\lambda)$ in field space satisfying 
\begin{equation} \label{construction}
\frac{d \Phi_n(\lambda)}{d \lambda} = F_n[\Phi_m].
\end{equation}
Notice that time-dependent solutions of the equations of motion, in general, do not follow such trajectories, and vice versa, only some of the $\Phi_n(\lambda)$ trajectories correspond to time-dependent solutions. 
For $N$ fields, the space of different trajectories is $(N-1)$-dimensional. That is, to encompass all trajectories generated by the symmetry, we need to introduce $(N-1)$ parameters on top of $\lambda$. We can conventionally assume that the $(N-1)$-dimensional submanifold that $\Phi_m(0)$ spans when we vary these parameters, contains our SSP solution $\bar \phi_n(0)$ evaluated at $t=0$.
The submanifold spanned by $\Phi_m(0)$ should be thought of as ``orthogonal" to the action of the symmetry generator, though in the absence of a natural metric in field space there is no unique definition or parameterization of it. If the symmetry group $G$ of the theory contains other generators beside that associated with our SSP solution (which we denote by $F$) part of the $\Phi_m(0)$ space can be reached by acting on $\bar \phi_n(0)$ with the coset $G/F$. In general, $\Phi_m(0)$ will contain all the heavy 
(``Higgs") degrees of freedom as well. Once the $\Phi_m(0)$ sub-manifold  is determined and parameterized according to some prescription, then the whole neighborhood we are interested in---the tube---is described by the $(N-1)$ parameters spanning $\Phi_m(0)$ and by  $\lambda$---the ``distance" from the $\Phi_m(0)$ manifold along the symmetry direction.

Fluctuations around an SSP solution are conveniently characterized in terms of the field-space coordinates defined above. First, we  pick an arbitrary parameterization of the  $\Phi_m(0)$ sub-manifold, possibly recovering the background solution  $\bar \phi_n(0)$ in the limit when the parameters---the `transverse' fluctuation fields---go to zero. Notice that such a manifold is, by definition, \emph{time-independent}; consequently, the specific choice  of coordinates we put on it is irrelevant for our purposes---finding a parameterization of field space that makes the fluctuation Lagrangian manifestly time-independent. Then we pose $\lambda = \pi + c t$. The time-dependent field transformation thus reads
\begin{equation} \label{ddefinition}
\phi_n = \Phi_n(\pi + c\, t) \, .
\end{equation} 
Since 
\begin{equation}
\frac{\partial \phi_n}{\partial t} = c \frac{d \Phi_n}{d \lambda} = c F_n
\end{equation}
by construction (eq.~\eqref{construction}), then the Lagrangian
\begin{equation}
{\cal L}_{\rm fluc}[\Phi_m(0), \pi,t] \equiv  {\cal L}[\Phi_n(\pi + c\, t)]
\end{equation}
is manifestly time-independent
\be
\frac{\partial {\cal L}_{\rm fluc}[\Phi_m(0), \pi,t]}{\partial t}  = 0  \; ,
\ee
in virtue of \eqref{symmetry!}. Since this holds exactly in this parameterization,  this is obviously true also at any order in perturbation theory about the SSP solution.

\subsection{Linearly realized symmetry}

As an application of the method described above, we now consider a linearly realized symmetry on a set of $N$ scalar fields, 
\begin{equation}
\delta \phi_n = F_n[\phi_m] = \tau_{nm} \, \phi_m\, .
\end{equation}
It is not restrictive to assume a real orthogonal (although possibly reducible) representation of this symmetry \cite{weinberg}. We therefore assume the generator $\tau_{nm}$ to be real and anti-symmetric. Note, moreover, that there could be other symmetries in the theory, and our $\tau$ could be one of the generators of a larger group. We will discuss the $SO(N)$ case in more detail shortly. 
The linearity of the representation allows us to write the SSP solutions and the solution of eq.~\eqref{construction} straightaway
(matrix and vector indices are omitted when possible): 
\begin{equation} \label{SSPlinear} 
\bar \phi(t) = U(ct) \cdot \bar\phi (0), \qquad \Phi(\lambda) = U(\lambda) \cdot \Phi(0) \; , 
\end{equation}
where $U$ is the orthogonal matrix
\begin{equation}
U(\lambda) = e^{\lambda \tau} \; .
\end{equation}

For definiteness, let us consider the simple Lagrangian 
\begin{equation}
{\cal L} = -\sfrac{1}{2} \partial_\mu \phi_m \partial^\mu \phi_m - V(\phi_m).
\end{equation}
By inserting the SSP condition \eqref{SSPlinear} into the equations  of motions we obtain a geometrical constraint on $\bar \phi_m$,
\begin{equation} \label{conditioN}
c^2 \, \tau^2 \cdot \bar\phi + \nabla V (\bar \phi)= 0.
\end{equation}
According to the  strategy outlined above, we parametrize the fluctuations as 
\begin{equation} \label{fluct linearly}
\phi = U(\pi + ct) \cdot \Phi \; ,
\end{equation}
which gives
\begin{equation} \label{L linearly}
{\cal L} = -\sfrac{1}{2} ( \partial_\mu (\pi + ct) \big)^2\big| \tau \cdot \Phi|^2 - \partial_\mu (\pi + ct) \, \partial^\mu \Phi \cdot \tau \cdot \Phi - \sfrac{1}{2} \big|\partial_\mu\Phi \big|^2 - V(\Phi) \; .
\end{equation}
The potential only depends on $\Phi$ because, by assumption,  it is invariant under the transformation  $U$. As advertised, with this parameterization of field space the fluctuation Lagrangian is manifestly time-independent---there is a derivative acting on every $t$.

We can be more specific in defining  the subspace spanned by $\Phi$. For instance, we can demand that $\Phi$ be the $(N-1)$-dimensional space orthogonal to the action of $\tau$ at $\bar \phi(0)$:
\begin{equation}
\bar \phi (0) \cdot  \tau \cdot \Phi = 0\, .
\end{equation}
Then, we can decompose the vectors in $\Phi_m$ into parallel and transverse to $\bar \phi(0)$:
\begin{equation} \label{sigma and v}
\Phi_n = \bar\phi_n(0) (1 + \sigma)  + v_n  \; , \qquad \bar \phi(0) \cdot v = 0, \qquad \bar \phi_n(0) \cdot \tau \cdot v  = 0\, .
\end{equation}
The field $\sigma$  parameterizes the ``radial mode'', which we expect to be massive like in ordinary SSB cases. The vector $v_n$ is a set of $(N-2)$ fields that is transverse both to $\bar \phi_n(0)$ and to its  variation under $\tau$.
With this parameterization of fluctuations, the unperturbed SSP solution is recovered for $\pi, v_n, \sigma \rightarrow 0$.


\subsection{The $SO(N)$ case}

Further simplifications are obtained by demanding that the Lagrangian be invariant under generic $SO(N)$ transformations on $\phi_n$, where $N$ is the number of fields.
In such a case the potential depends only on the norm of $\phi_n$ and, by \eqref{conditioN}, $\tau^2 \cdot \bar \phi(0)$ is thus parallel to $\bar \phi(0)$ itself. We can then normalize our $\tau$ so that
\begin{equation}\label{so(n)}
\tau^2 \cdot \bar\phi  = -  \bar\phi   \;  ,
\end{equation}
where the minus sign is necessary because $\tau$ is real and antisymmetric: $i \cdot \tau$ is hermitian, thus diagonalizable with real eigenvalues. Its square, which is $- \tau^2$, is a non-negative matrix.

In the presence of the full $SO(N)$ group, there is a more convenient parameterization than eq.~\eqref{fluct linearly} that puts the kinetic terms for $v $ directly in canonical form. It amounts to defining
\be
\phi = U(ct + \pi) \cdot \bar \phi(0) (1+ \sigma) + v' \; ,
\ee
or equivalently, to redefine $v$ as
\be
v = U(-(ct+\pi)) \cdot v' \; .
\ee
This is also equivalent to define last subsection's $U$ as the exponential not of our $\tau$, but of a linear combination of $\tau$ and of other $SO(N)$ generators that acts as $\tau$ itself on the SSP  solution $\bar \phi(t)$, but that acts trivially on $v$ \footnote{
Because of \eqref{so(n)}, the SSP solution $U(ct) \cdot \bar \phi(0)$ describes a circle in field space. The $v_n$'s parameterize the $N-2$ dimensional subspace orthogonal to this circle. In the plane of the circle, $\tau$ acts as the $2D$ rotation generator
\be
\begin{pmatrix} 
0 & 1 \\
-1& 0\\
\end{pmatrix} \; .
\ee
We can set all other entries to zero by taking suitable linear combinations with other generators that act trivially in the plane of the circle. Indeed, a convenient basis for the $SO(N)$ generators is
\be
\tau^{(12)} \equiv
\begin{pmatrix} 
0 & 1& 0 & \dotsi \\
-1& 0& 0 & \dotsi\\
0& 0 & 0& \dotsi \\
\vdots& \vdots &\vdots &\ddots	
\end{pmatrix} \; ,
\qquad \mbox{etc.}
\ee
}.  We need these other $SO(N)$ generators to correspond to actual  symmetries of the  Lagrangian in order not to impair the arguments of
sect.~\ref{sec.5.1}.

With this new parameterization for fluctuations, the Lagrangian reduces to (dropping the primes)
\begin{eqnarray} \nonumber
{\cal L}& = & -\sfrac{1}{2} \, (f+\sigma)^2 \, (\partial \pi)^2 - \sfrac{1}{2} (\partial \sigma)^2 -  \sfrac{1}{2} |\partial v|^2  + c \, \dot \pi (f+\sigma)^2 \nonumber\\ 
&&     + \sfrac{1}{2} c^2 (f+\sigma)^2   - V(|\Phi|) \label{Lagrr} \; ,
\end{eqnarray}
where we defined the symmetry breaking scale $f$ as
\be
f^2 = \big| \bar \phi(0) \big |^2 \; ,
\ee
and we redefined $\sigma$ by absorbing a power of $f$ into it.
We have collected all the non-derivative terms in the second line.

In standard SSB, there are as many massless Goldstone bosons as the number of broken generators. Our $N$-vector $\phi_n$ spontaneously breaks $SO(N)$ down to $SO(N-1)$, that is, it breaks $N-1$ generators. In our parameterization, the corresponding $N-1$ Goldstone bosons would be $\pi$ and all the $v_n$'s. 
However, while it is clear from the above Lagrangian that $\pi$ is a flat direction---it is always acted upon by derivatives---for the $v_n$'s we have  mass terms coming from the potential $V$ itself, since $|\Phi|$ depends on $v$:
\be
|\Phi| = \sqrt{(f+\sigma)^2 + |v|^2} \; .
\ee
At quadratic order in fluctuations one is left simply with
\begin{eqnarray}
{\cal L}& = & -\sfrac12 {f^2} (\partial \pi)^2 - \sfrac{1}{2} (\partial \sigma)^2 -  \sfrac{1}{2} |\partial v|^2 + 2 c f  \, \dot \pi \sigma \\
&& -\sfrac{1}{2} (V''(f) - c^2 )  \sigma^2 - \sfrac{1}{2} c^2 |v|^2 + \dots \; ,
\end{eqnarray}
where we used the eom for the background eq.~\eqref{conditioN}, which is here equivalent to canceling the tadpole for $\sigma$:
\be
c^2  f = V'(f) \; . 
\ee
As predicted, in the presence of  SSP the $(N-2)$ ``non-rotating" Goldstones get a mass of order $c$.

\section{Concluding remarks: Application to cosmology}

Models of primordial inflation offer an ideal application of our formalism. In most cases, the observed near scale-invariance of primordial density perturbations is a consequence of the inflationary phase's approximate de Sitter isometries, which are in turn a consequence of an approximate internal symmetry of the inflaton Lagrangian---typically a shift symmetry. The inflaton time-dependent background solution {\em spontaneously probes} this approximate symmetry.

Symmetry considerations are also heavily used in the recently introduced
effective field theory for adiabatic inflationary perturbations---the so-called effective field theory of inflation \cite{CCFKS}. There, one thinks of adiabatic perturbations as the Goldstone bosons of spontaneously broken time-translations
\footnote{In a theory with dynamical gravity space-time translations are gauged. As a consequence, the would-be Goldstone bosons are `eaten' by the gravitational degrees of freedom---there is a gauge in which the Goldstones are set to zero, and the metric has propagating longitudinal degrees of freedom on top of the usual helicity-2, transverse ones. The Goldstone boson language is still useful though: like for massive gauge theories, at short distances the dynamics of these extra gravitational degrees of freedom are correctly captured by those of the Goldstones.}
\cite{CLNS}.
This characterization has far-reaching implications for their dynamics, and for the observationally relevant quantities---the correlation functions of density perturbations. What this approach still lacks, however, is a way of systematically dealing with the (approximate) {\em internal} symmetries that are spontaneosly broken during inflation. For instance, the implications of the assumed approximate shift symmetry of the inflaton Lagrangian, and of its spontaneous breaking, are never taken beyond the zeroth-order statement that the coefficients of the perturbations' effective Lagrangian are nearly constant in time. However it is not clear in general what their mild time-dependence looks like, nor what the analogous zeroth-order statement is in more general situations. Consider for instance a scalar field theory with {\em exact} shift-invariance---e.g.~a $P(X)$-theory---driving an FRW cosmology that is {\em not} an approximately deSitter inflation phase. For instance, a free theory with ${\cal L} =- \sfrac12 (\di \phi)^2$, admits a cosmological solution with $\dot \phi \sim 1/a^3$, $a \sim t^{1/3}$. One can still apply the EFT construction of \cite{CLNS, CCFKS} to figure out the general implications of spontaneously broken time-translations for the perturbations' dynamics. However, now nothing depends {\em mildly} on time, and it is not obvious what constraints are imposed on the perturbations' time-dependent Lagrangian coefficients by the spontaneously broken internal shift-symmetry.

Our formalism---which explicitly addresses the simultaneous breakdown of time-translations and of internal symmetries---is a first step to correct these shortcomings. In order for it to be a useful addition to the effective field theory of adiabatic perturbations however, we need first of all to include gravity in our analysis. Moreover, in most models of inflation the inflaton shift symmetry is only approximate---if it were exact,  inflation could not end
\footnote{A notable exception is ghost inflation \cite{ghost inflation}.}. 
Therefore, we need to include in our analysis the effects of weak explicit symmetry breaking terms as well. We carry out this generalization of our results to inflationary cosmology in a forthcoming publication.

\section*{Acknowledgments}

We thank Sarah Shandera for many useful discussions. 
The work of AN is supported by the DOE under contract DE-FG02-11ER1141743 and by NASA under contract NNX10AH14G.

\appendix

\section*{Appendix}

\section{The running Goldstone field beyond linear order}

In Sec. \ref{sec4.2} we defined the Goldstone field $\pi(x)$ as the linear piece of $J^\mu$ (eq.~\eqref{pi}), by exploiting the property that the current interpolates the running Goldstone states. This defines the $\pi(x)$ field operator up to non-linear field redefinitions of the form
\begin{equation} \label{transformtilda}
\tilde \pi(x) = \pi(x) + {\cal O}(\pi^2),
\end{equation}
We now show that we can consistently impose the transformation property 
\eqref{transformQ} and that this fixes the residual ambiguity in the definition of $\pi(x)$. 

According to \eqref{junk}, for a canonically normalized Goldstone field we have
\begin{equation}\label{appendidi}
\delta_Q \pi(x) = v + G[\pi] \; ,
\end{equation}
where $G$ is a local functional that starts linear in $\pi$. 
The general treatment of non-linearly realized symmetries---see e.g.~\cite{CWZ}---typically assumes that internal symmetries act locally on the field's value at any given $x$, with no dependence on the field's values at nearby points. That is, it assumes that $G$ above is a local {\em function} of $\pi$,
\be \label{local G}
G[\pi] \to G\big(\pi(x) \big ) \; ,
\ee
with no dependence on $\pi$'s derivatives. Notice that such an assumption is somewhat restrictive---for instance, it can always be evaded by performing a non linear field redefinition that {\em does} depend on derivatives, e.g.
\be
\pi (x) = \pi' + \alpha (\di \pi')^2 \; .
\ee
However, in  low-energy effective field theory one usually organizes the theory---the action and its symmetries---as an expansion in derivatives. The lowest order dynamics in this expansion are invariant under the zero-derivative reduction of the symmetries. According to  this viewpoint, at least at lowest order in the derivative expansion we can stick to  the no-derivative version of $G[\pi]$, eq.~\eqref{local G}, which we get by simply ignoring all derivatives possibly appearing in \eqref{appendidi}. 
In such a case we just have
\be
\delta_Q \pi = v + G(\pi) \; ,
\ee
whereas we want
\be
\delta_Q \pi' = v \; .
\ee
The field redefinition accomplishing this is obviously
\be
\pi' (x)=  v \int ^{\pi(x)} _0 \frac{d \pi}{v + G(\pi) }  \; .
\ee
The question of whether there exist genuine derivative corrections to \eqref{appendidi}---genuine in the sense that they cannot be removed by a field redefinition involving derivatives---is an interesting one, which for the time being we have  nothing to say about.

It is interesting to look at things from a different perspective. At the operator level, eq.~\eqref{appendidi} reads
\be
 i \int d^3 y [J^0(y),\pi(\vec x)] = 1 + G[\pi] \; ,
\ee
Asking for a field redefinition that eliminates $G$, is equivalent to asking for a parameterization of field space such that $J^0(x) = \Pi(x)$, where $\Pi(x)$ is the momentum conjugate to $\pi$. In the semiclassical case considered in sect.~\ref{sec.5}, if the Lagrangian does not contain more than one time derivative on each field this condition reads\footnote{The functionals $F_n$ are defined in \eqref{transformationss} as the generators of the symmetry transformation.  In general, the definition of the current also contains the functional $K^\mu$, whose divergence gives the variation of the Lagrangian under the symmetry. However, for internal symmetries, we know of no cases where $K^\mu$ cannot be set to zero by adding suitable total derivatives to the Lagrangian.}
\begin{equation}
\frac{\partial {\cal L}}{\partial \dot \phi_n} F_n \, = \, \frac{\partial {\cal L}}{\partial \dot \pi} \, .
\end{equation} 
This is accomplished precisely by the field-space parameterization discussed in Sec.~\ref{sec.5.1}. Indeed, from eqs.~\eqref{construction} and \eqref{ddefinition} we get
\begin{equation}
\dot \phi_n = \dot \Phi_n+ \big(\dot \pi + c \big) \cdot F_n \; .
\end{equation}
With this parameterization eq. \eqref{appendidi}, with $G=0$,  follows.



%

\section{Absence of extra contact terms in eq.~\eqref{T with pi}}

The Ward identity associated with time-translational invariance  reads:
\begin{align}
\di^{(x)}_\mu \langle  \, T^{\mu0}(x) \, \pi(x_1) \dots \pi(x_n) \, \rangle_T &  = 
-i \delta^4(x-x_1) \, \langle \, \delta_H \pi(x_1) \dots \pi(x_n) \, \rangle_T  - \\
& \dots  -i \delta^4(x-x_n) \, \langle \,  \pi(x_1) \dots \delta_H \pi(x_n) \, \rangle_T \; ,
\end{align}
where the expectation value is taken on our SSP vacuum $| c \ket$, and the subscript $T$ denotes $T$-ordering. The local form of the Ward identity 
is exact even though time-translations are spontaneously broken by $|c \ket$. We can bring the derivative inside the $T$-ordered product, but this generates extra terms coming from hitting the $\theta$-functions in $T(\dots)$ with the time derivative. We get 
\begin{align} \label{Ward}
\langle  \,   \di_\mu T^{\mu0}(x) \, & \pi(x_1) \dots \pi(x_n)  \, \rangle_T  + \delta(t-t_1) \langle  \, \big[ T^{00}(x) , \pi(x_1) \big]  \dots \pi(x_n) \,  \rangle_T  \nonumber \\
&  + \dots  + \delta(t-t_n) \langle  \,\pi(x_1) \dots  \big[ T^{00}(x) , \pi(x_n) \big] \,  \rangle_T \nonumber \\
&  = -i \delta^4(x-x_1) \, \langle \, \delta_H \pi(x_1) \dots \pi(x_n) \, \rangle_T  - 
 \dots  -i \delta^4(x-x_n) \, \langle \,  \pi(x_1) \dots \delta_H \pi(x_n) \, \rangle_T \; .
\end{align}
The operator ${\cal O} (x)\equiv \di_\mu T^{\mu0}(x)$ vanishes identically. To see this, in \eqref{Ward}  we can take the time of ${\cal O}$ to be different from the times of all the $\pi$'s. All $\delta$-functions then vanish and  we get
\be
\langle  {\cal O} \, \pi \dots \pi  \rangle _T = 0 \; ,
\ee
for an arbitrary number of $\pi$ fields.
All (unequal-time) correlation functions of ${\cal O}(x)$ vanish, which means that ${\cal O}(x)$ is a trivial operator. 
If we then go back  to \eqref{Ward} and now take all $t_1, \dots, t_n$ to be different from each other, we are left with $n$ independent equations, of the form
\be
\delta(t-t_1) \langle  \, \big[ T^{00}(x) , \pi(x_1) \big]  \dots \pi(x_n) \,  \rangle_T   = -i \delta^4(x-x_1) \, \langle \, \delta_H \pi(x_1) \dots \pi(x_n) \, \rangle_T  
\ee
Or, dropping the common $\delta(t-t_1) $ factor from both sides, and setting $t_1=t$:
\be
\langle  \, \big[ T^{00}(\vec x, t) , \pi(\vec x_1, t) \big]  \dots \pi(x_n) \,  \rangle_T   = -i \delta^3(\vec x- \vec x_1) \, \langle \, \delta_H \pi(\vec x_1, t) \dots \pi(x_n) \, \rangle_T \; .
\ee
That is, the equal time  $[T^{00}, \pi]$ commutator has (unequal-time) correlation functions with an arbitrary number of $\pi$ fields as dictated by  eq.~\eqref{T with pi}. It means that eq.~\eqref{T with pi} has to hold at the operator level, without further contact terms.


\begin{thebibliography}{99}

\bibitem{ghost}
  N.~Arkani-Hamed, H.~-C.~Cheng, M.~A.~Luty, S.~Mukohyama,
 ``Ghost condensation and a consistent infrared modification of gravity,''
  JHEP {\bf 0405}, 074 (2004).
  [hep-th/0312099].

\bibitem{ghost2}
  N.~Arkani-Hamed, H.~-C.~Cheng, M.~A.~Luty, S.~Mukohyama, T.~Wiseman,
  ``Dynamics of gravity in a Higgs phase,''
  JHEP {\bf 0701}, 036 (2007).
  [hep-ph/0507120].

  \bibitem{WB}
  H.~Watanabe, T.~Brauner,
  ``On the number of Nambu-Goldstone bosons and its relation to charge densities,''
  [arXiv:1109.6327 [hep-ph]].

\bibitem{coleman}
S.~Coleman, 
\emph{Aspects of Symmetry, Selected Erice Lectures}, 
Cambridge University Press, 1985, p.~258.

\bibitem{son}
D.~T.~Son,
``Low-Energy Quantum Effective Action for Relativistic Superfluids,''
arXiv:hep-ph/0204199.
 
\bibitem{NRT}
A.~Nicolis, R.~Rattazzi and E.~Trincherini,
``Energy's and Amplitudes' Positivity,''
JHEP {\bf 1005} (2010) 095
[Erratum-ibid.\ {\bf 1111} (2011) 128]
[arXiv:0912.4258 [hep-th]].

\bibitem{dRT}
C.~de Rham and A.~J.~Tolley,
``DBI and the Galileon Reunited,''
JCAP {\bf 1005} (2010) 015
[arXiv:1003.5917 [hep-th]].

\bibitem{rubakov}
  V.~A.~Rubakov,
  ``Harrison-Zeldovich spectrum from conformal invariance,''
  JCAP {\bf 0909}, 030 (2009).
  [arXiv:0906.3693 [hep-th]].
  
\bibitem{justin}
  K.~Hinterbichler, J.~Khoury,
  ``The Pseudo-Conformal Universe: Scale Invariance from Spontaneous Breaking of Conformal Symmetry,''

\bibitem{dbi}
  M.~Alishahiha, E.~Silverstein, D.~Tong,
  ``DBI in the sky,''
  Phys.\ Rev.\  {\bf D70}, 123505 (2004).
  [hep-th/0404084].


\bibitem{genesis}
P.~Creminelli, A.~Nicolis and E.~Trincherini,
``Galilean Genesis: an Alternative to Inflation,''
JCAP {\bf 1011} (2010) 021
[arXiv:1007.0027 [hep-th]].



\bibitem{NP}
A.~Nicolis and F.~Piazza,
``A relativistic non-relativistic Goldstone theorem",
in preparation.

  
  


  

%


\bibitem{CN}
P.~Creminelli, A.~Nicolis,
``On the regime of validity of derivatively coupled theories'',
in preparation.

\bibitem{NR}
A.~Nicolis and R.~Rattazzi,
``Classical and Quantum Consistency of the DGP Model,''
JHEP {\bf 0406} (2004) 059
[arXiv:hep-th/0404159].

\bibitem{Goldstone}
  J.~Goldstone, A.~Salam, S.~Weinberg,
  ``Broken Symmetries,''
  Phys.\ Rev.\  {\bf 127}, 965-970 (1962).

\bibitem{landau}
E.~M.~Lifshitz and L.~P.~Pitaevskii,
``Statistical physics. Part 2: Theory of the condensed state,''
{\em Oxford, UK: Butterworth-Heinemann (1980) 387p.}

\bibitem{kibble}
  G.~S.~Guralnik, C.~R.~Hagen, T.~W.~B.~Kibble,
  ``Global Conservation Laws and Massless Particles,''
  Phys.\ Rev.\ Lett.\  {\bf 13}, 585-587 (1964).

\bibitem{NC}
H.~B.~Nielsen and S.~Chadha,
``On How to Count Goldstone Bosons,''
Nucl.\ Phys.\ B {\bf 105} (1976) 445.

\bibitem{lange}
  R.~V.~Lange,
  ``Goldstone Theorem in Nonrelativistic Theories,''
  Phys.\ Rev.\ Lett.\  {\bf 14}, 3-6 (1965).

\bibitem{brauner}
  T.~Brauner,
  ``Spontaneous Symmetry Breaking and Nambu-Goldstone Bosons in Quantum Many-Body Systems,''
  Symmetry {\bf 2}, 609-657 (2010).
  [arXiv:1001.5212 [hep-th]].
  
  
\bibitem{weinberg}
  S.~Weinberg,
  ``The quantum theory of fields. Vol. 2: Modern applications,''
  Cambridge, UK: Univ. Pr. (1996) 489 p.




\bibitem{CCFKS}
C.~Cheung, P.~Creminelli, A.~L.~Fitzpatrick, J.~Kaplan and L.~Senatore,
``The Effective Field Theory of Inflation,''
JHEP {\bf 0803} (2008) 014
[arXiv:0709.0293 [hep-th]].

\bibitem{CLNS}
P.~Creminelli, M.~A.~Luty, A.~Nicolis and L.~Senatore,
``Starting the Universe: Stable Violation of the Null Energy Condition and Non-Standard Cosmologies,''
JHEP {\bf 0612} (2006) 080
[arXiv:hep-th/0606090].


\bibitem{ghost inflation}
N.~Arkani-Hamed, P.~Creminelli, S.~Mukohyama and M.~Zaldarriaga,
``Ghost Inflation,''
JCAP {\bf 0404} (2004) 001
[arXiv:hep-th/0312100].

\bibitem{CWZ}
S.~R.~Coleman, J.~Wess and B.~Zumino,
``Structure of Phenomenological Lagrangians. 1,''
Phys.\ Rev.\ {\bf 177} (1969) 2239.





\end{thebibliography}
\end{document}